\documentclass[journal=nalefd,manuscript=letter]{achemso}

\usepackage{graphicx}
\usepackage[ansinew]{inputenc}
\usepackage{amsmath,amsfonts,amssymb,mathtools}
\usepackage{color}

\title{Understanding and engineering phonon-mediated tunneling into graphene on metal surfaces}

\author{J. Halle}
\email{johannes.halle@tu-ilmenau.de}
\affiliation{Institut für Physik, Technische Universität Ilmenau, D-98693 Ilmenau, Germany}
\author{N. Néel}
\affiliation{Institut für Physik, Technische Universität Ilmenau, D-98693 Ilmenau, Germany}
\author{M. Fonin}
\affiliation{Fachbereich Physik, Universität Konstanz, D-78457 Konstanz, Germany}
\author{M. Brandbyge}
\affiliation{Center for Nanostructured Graphene, Department of Micro- and Nanotechnology, Technical University of Denmark, DK-2800 Kongens Lyngby, Denmark}
\author{J. Kröger}
\affiliation{Institut für Physik, Technische Universität Ilmenau, D-98693 Ilmenau, Germany}


\begin{document}
\newpage

\begin{abstract}
Metal-intercalated graphene on Ir(111) exhibits phonon signatures in inelastic electron tunneling spectroscopy with strengths that depend on the intercalant.
Extraordinarily strong graphene phonon signals are observed for Cs intercalation.  Li intercalation likewise induces clearly discriminable phonon signatures, albeit less pronounced than observed for Cs. 
The signal can be finely tuned by the alkali metal coverage and gradually disappears upon increasing the junction conductance from tunneling to contact ranges.
In contrast to Cs and Li, for Ni-intercalated graphene the phonon signals stay below the detection limit in all transport ranges.
Going beyond the conventional two-terminal approach, transport calculations provide a comprehensive understanding of the subtle interplay between the graphene--electrode coupling and the observation of graphene phonon spectroscopic signatures.

\begin{tocentry}
\begin{center}
\includegraphics[width=0.8\textwidth]{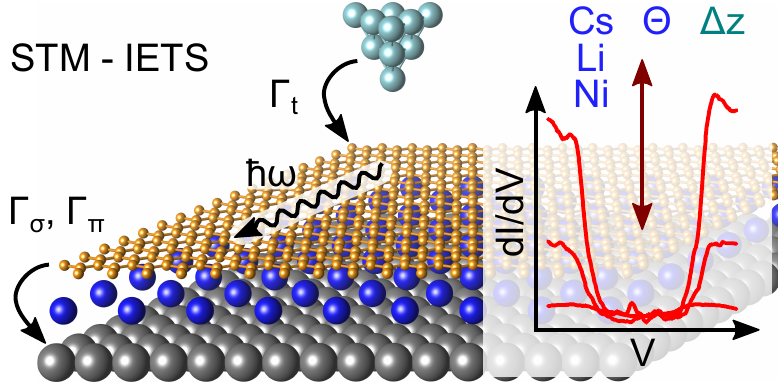}
\end{center}
\label{TOC}
\end{tocentry}

\end{abstract}

\maketitle

Graphene phonons are relevant to technological applications and fundamental research alike.
For instance, the scattering of electrons with optical graphene phonons affects the electron transport properties of graphene in the high-current limit \cite{prl_103_076601}.
In addition, the relation between phonons and thermal transport in graphene attracts increasing attention \cite{rpp_80_036502}.
Furthermore, the electron--phonon coupling strength \cite{prl_93_185503}, the possible distortion of the Dirac cone \cite{nl_10_4335}, and the graphene--substrate hybridization \cite{njp_15_043031} may be inferred from the inspection of phonon dynamics.
Local probes of graphene phonons are particularly appealing since they enable the examination of the influence of adsorbates, defect sites, doping and the graphene--substrate interaction on the C lattice vibrations at the atomic scale.
Inelastic electron tunneling spectroscopy (IETS) with a scanning tunneling microscope (STM) has so far been used to explore phonons of graphene on semiconducting or nearly insulating substrates including SiC \cite{apl_91_122102}, SiO$_2$ \cite{natphys_4_627,prl_104_036805}, and SiO$_2$ covered with hexagonal boron-nitride \cite{nl_11_2291,prl_114_245502}. Recently, IET signals of graphene phonons have been reported from delaminated graphene nanostructures on Pt(111) \cite{natcommun_6_7528} and Ir(111) \cite{prb_95_161410} as well as from bilayer graphene on Ir(111) \cite{small_14_1703701}.
 
At present, the occurrence of graphene phonon signals in IETS is far from being understood.
It seems that nearly free graphene, \textit{i.\,e.}, a weak graphene--substrate hybridization, favors the conservation of the genuine graphene electronic structure and the concomitant phonon-mediated tunneling \cite{apl_91_122102,natphys_4_627,prl_104_036805,nl_11_2291,prl_114_245502,small_14_1703701,natcommun_6_7528,prb_95_161410}.
However, in some tunneling spectroscopy studies of exfoliated graphene on SiO$_2$ phonon signatures were not observed \cite{prb_79_205411,natphys_7_245}.
Moreover, so far experiments and simulations have solely considered this weak hybridization limit and the coupling between graphene and adjacent electrodes has not been explicitly modeled to date.
Therefore, the relation between the signal strength of graphene phonon signatures in IETS and the graphene--electrode coupling remains elusive.

Here, we present a combination of IETS experiments and transport calculations, which unambiguously unveils the intimate relation between the covalent graphene--electrode coupling and the IET signal strength of graphene phonons. Details on sample preparation and experimental methods can be found in the Supporting Information (Section 1).
In contrast to previous work \cite{apl_91_122102,natphys_4_627,prl_104_036805,nl_11_2291,prl_114_245502,small_14_1703701,natcommun_6_7528,prb_95_161410}, graphene-covered Ir(111) intercalated by Cs, Li, Ni represents an all-metal complex in which the graphene--substrate interaction is tailored by the chemical nature and the amount of the intercalant.
In addition, the tip--graphene hybridization is finely tuned by controllably changing the tip--graphene separation from tunneling to contact distances.
The observed different IET signals of graphene phonons are not in agreement with the expected trend with the charge carrier density \cite{natphys_4_627,prl_104_036805,nl_11_2291,prl_114_245502}. Our data provide the basis for developing a general picture of inelastic electron transport across graphene on surfaces.
Transport calculations based on density functional theory (DFT) using a multi-electrode setup enable the analysis of the branching of the electron current from the STM tip into graphene and the substrate. 
The DFT findings are translated into a simplified model that provides an intuitive understanding of the relation between the graphene--electrode hybridization and the effective phonon excitation.

\begin{figure}
\includegraphics{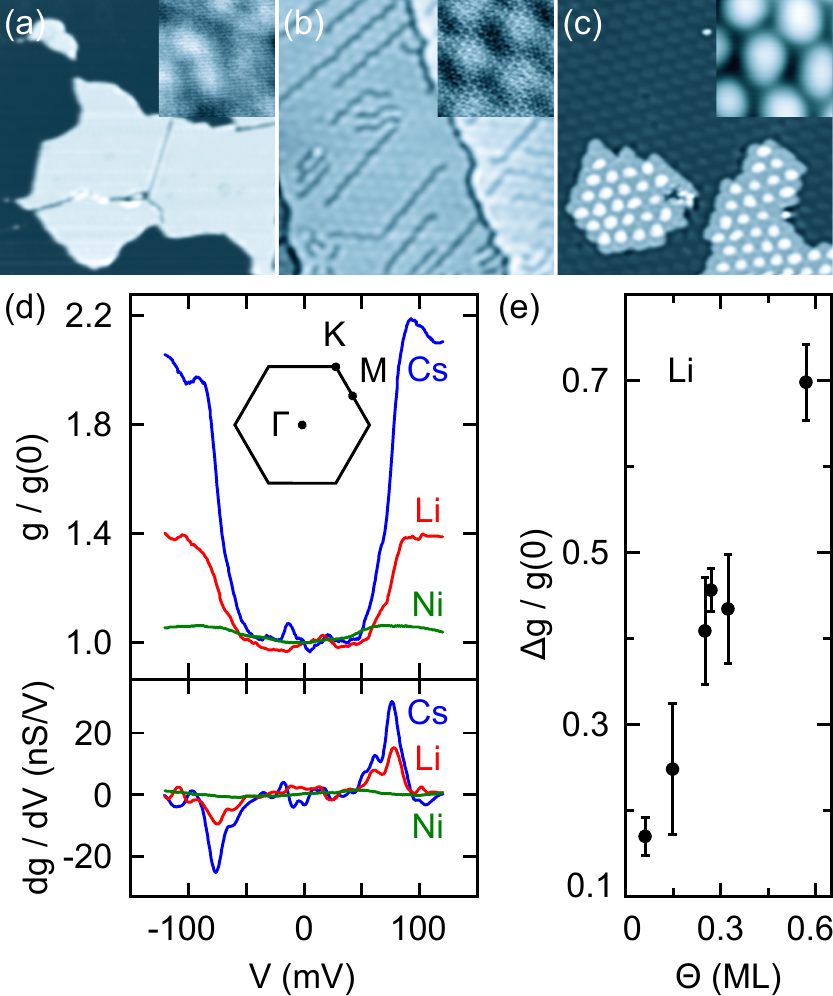}
\caption{(Color online) 
Constant-current STM images (tunneling current $I=100\,\text{pA}$, bias voltage $V=120\,\text{mV}$ applied to the sample) of graphene-covered Ir(111) intercalated by (a) Cs ($100\times 100\,\text{nm}^2$), (b) Li ($50\times 50\,\text{nm}^2$), (c) Ni ($40\times 40\,\text{nm}^2$)\@. 
Insets to (a)--(c): Close-up views ($5.5\times 5.5\,\text{nm}^2$)\@.  
The gray scale ranges from $0\,\text{pm}$ to (a) $10\,\text{pm}$, (b) $30\,\text{pm}$, (c) $150\,\text{pm}$.
In the insets to (b), (c) graphene moiré patterns are visible.
STM data were processed with WSxM \cite{rsi_78_013705}.
(d) Top panel: $g = \text{d}I/\text{d}V$ spectra of the intercalated samples, normalized to the zero-bias differential conductance, $g(0)$\@. 
Cs and Li data exhibit steplike signatures that are ascribed to the excitation of graphene phonons, while Ni data are essentially featureless. 
Inset: Surface Brillouin zone of graphene with indicated high-symmetry points.
Bottom panel: Numerical derivative ($\text{d}g/\text{d}V$) of the spectra in the top panel.
(e) Phonon-induced change in $\text{d}I/\text{d}V$ ($\Delta g$) divided by the zero-bias differential conductance $g(0)$ as a function of the Li coverage $\Theta$\@.
\label{fig1}
}
\end{figure}

Figure \ref{fig1} shows STM images of graphene-covered Ir(111) intercalated by Cs (Figure~\ref{fig1}a), Li (Figure~\ref{fig1}b), Ni (Figure~\ref{fig1}c)\@.
In all cases the intercalated metal film exhibits monatomic height. 
With respect to graphene, Cs and Li intercalate with a, respectively, $(2\times2)$ and $(\sqrt{3}\times\sqrt{3})~\text{R30}^\circ$  superstructure.\cite{natcommun_4_2772, jpcc_120_5067}
For Ni intercalation, experiments indicated a pseudomorphic growth on Ir(111)  \cite{prb_87_035420}.
The insets to the STM images reveal that the moiré superstructure of pristine graphene remained after intercalation to different extents.
While the corrugation of Cs-intercalated graphene is below the resolution limit, Li-intercalated (Ni-intercalated) graphene exhibits a corrugation of $9\pm 1\,\text{pm}$ ($105\pm 3\,\text{pm}$)\@.
The moiré-induced corrugation of pristine graphene on Ir(111) at the same tunneling parameters is $19\pm 1\,\text{pm}$ \cite{jpcc_120_5067}.
Previously, the moiré corrugation was identified as a measure of the graphene--substrate hybridization \cite{prb_87_035420,prb_78_073401}.
Therefore, graphene on Cs may be characterized as well decoupled, shows a weak coupling for intercalated Li and is strongly hybridized with the Ni film.

For these intercalated samples IET spectra were recorded, which represent the main experimental finding of this work.
Figure \ref{fig1}d shows that Cs-intercalated graphene displays a gap-like feature, symmetrically positioned around zero bias.
Abrupt increases of $g=\text{d}I/\text{d}V$ occur at $\pm 56\,\text{mV}$ and $\pm 75\,\text{mV}$, which give rise to an enhancement of $g$ with respect to $\text{d}I/\text{d}V$ at zero bias, $g(0)$, exceeding $200\,\%$\@.
In accordance with previous results reported for graphene wrinkles \cite{prb_95_161410} and with the graphene phonon dispersion on Ir(111) \cite{prb_88_205403,annphys_526_372} these changes are assigned to out-of-plane acoustic ($\pm 56\,\text{mV}$), optical ($\pm 75\,\text{mV}$), and transverse acoustic ($\pm 56\,\text{mV}$) graphene phonons at the $M$ point of the surface Brillouin zone.
The same phonon spectroscopic signatures are visible for Li-intercalated graphene, albeit to a smaller extent; that is, $g$ is increased to $\approx 140\,\%$ of $g(0)$ at a Li coverage of $\Theta=0.27\,\text{ML}$, where $1\,\text{ML}$ (ML: monolayer) is defined by $1$ intercalant atom per C ring. Additional spectroscopic data for Li are presented in the Supporting Information (Figure~S1).
Ni-intercalated graphene does not reveal discernible variations in $\text{d}I/\text{d}V$ spectra due to phonon excitation.
For all intercalants a dependence of the spectra on the graphene position was not discernible (Supporting Information, Figure S2)\@.

Before entering into the discussion of the calculated results, it is worth mentioning that the phonon-induced changes in $\text{d}I/\text{d}V$ may be controlled to some extent by the coverage of the intercalants.
For Li we found that in the low submonolayer range relative changes, $\Delta g/g(0)$, are $\approx 17\,\%$ and increase up to $\approx 70\,\%$ for the densely packed Li film (Figure~\ref{fig1}e)\@. The formation of compact Cs islands even at low coverage hampered similar measurements for Cs-intercalated graphene.

\begin{figure}
\includegraphics{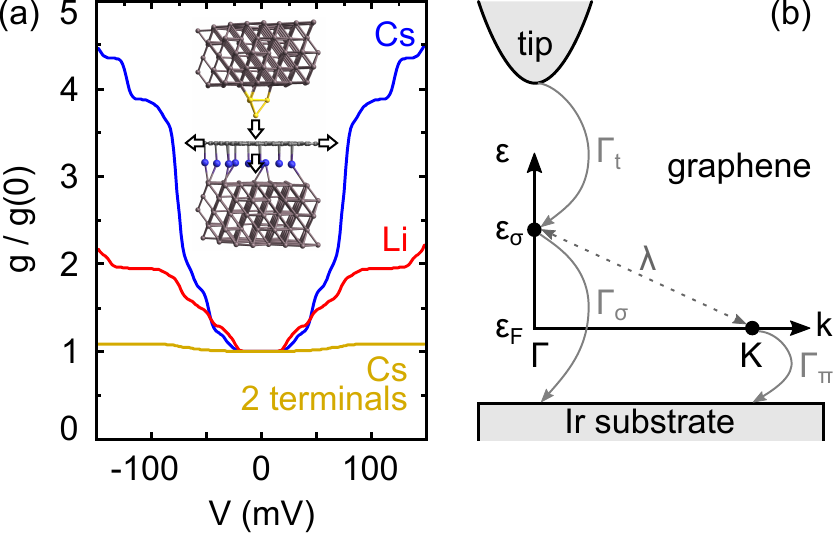}
\caption{(Color online) 
(a) Calculated $g/g(0)$ for graphene on Ir(111) intercalated by $(2\times2)~\text{Cs}$ and $(\sqrt{3}\times\sqrt{3})~\text{R30}^\circ~\text{Li}$ superlattices in the three-terminal setup.
In the conventional two-terminal approach (calculated for Cs, bottom curve) IET signals are virtually absent.
Inset: Setup for the calculations indicating (arrows) the presence of three terminals that collect propagating electrons in the tip, substrate and graphene.
(b) Illustration of electron transport in the three-terminal setup.
The tip couples to graphene $\sigma$ states with energy $\varepsilon_\sigma$ at $\Gamma$ with a coupling constant $\Gamma_\text{t}$\@.
The hybridization of graphene to the substrate is mediated by $\sigma$ states with strength $\Gamma_\sigma$ and by $\pi$ states with strength $\Gamma_\pi$. $\pi$ states occur at $K$ with energy $\varepsilon_\text{F}$ (Fermi energy). $\sigma$ and $\pi$ states are coupled by the electron-phonon interaction $\lambda$\@.
}
\label{fig2}
\end{figure}

Transport calculations (Supporting Information, Section~2) were performed in order to thoroughly understand the experimental results and to pinpoint the role of the graphene--substrate as well as the graphene--tip coupling in the IET signal strength for graphene phonons. In the following, tip and substrate will often be referred to as electrodes for simplicity.
The inset to Figure~\ref{fig2}a illustrates the setup for the calculations. 
Remarkably, a standard calculation with $\Gamma$-point approximation including two terminals --- tip and substrate --- and periodic boundary conditions in the transverse directions yields vanishing inelastic signatures in the current, even for Cs (Figure~\ref{fig2}a, bottom data set) and does not reproduce the experimental data.
However, introducing graphene self-energies, which is equivalent to attaching a third terminal that collects electrons propagating in graphene alone (Supporting Information, Figure~S3) results in a substantial enhancement of the phonon signatures. 
This setup considers the branching of the current into the metal substrate and graphene.
As shown in Figure~\ref{fig2}a this three-terminal model can qualitatively reproduce the experimental findings for $(2\times 2)~\text{Cs}$ and $(\sqrt{3}\times\sqrt{3})~\text{R30}^\circ~\text{Li}$. 
Quantitatively, the same order of magnitude for the phonon-induced changes in $g/g(0)$ is calculated, although they exceed the experimental values. 
In the calculations, the contributing phonon modes are similar to out-of-plane bands at $M$ and $K$, but the breaking of symmetry by the substrate also yields contributions shifted away from these (Supporting Information, Figure~S4)\@.

The findings based on density functional and transport calculations can be illustrated in a simple two-level model (Supporting Information, Section~2) involving the first unoccupied band $\sigma$ of graphene with energy $\varepsilon_\sigma$ at $\Gamma$ and a graphene $\pi$ state with energy $\varepsilon_\text{F}$ (Fermi energy) at $K$ (Figure~\ref{fig2}b)\@.
This model is inspired by previous work \cite{prl_101_216803}. 
The coupling of these states to the metal substrate is modeled by inverse lifetimes, $\Gamma_\sigma$ and $\Gamma_\pi$, where $\Gamma_\sigma>\Gamma_\pi$ due to the long range of $\sigma$ \cite{arxiv_1803_01568}.
Additionally, the $\sigma$ state is coupled to the tip with $\Gamma_\text{t}$\@.
Electrons injected from the tip into $\sigma$ can either directly continue to the substrate, which constitutes the elastic transport channel, or take the detour \textit{via} $\pi$ through electron--phonon coupling with strength $\lambda$.
In this inelastic transport channel a phonon with energy $\hbar\Omega$ is excited.
For $\hbar\Omega\ll\Gamma_\pi$, in the lowest-order expansion of the electron--phonon coupling and in the wide-band approximation \cite{prb_89_081405, nl_6_258} the relative conductance increase due to phonon excitation can be expressed as 
\begin{equation}
\frac{\Delta g}{g(0)}=\frac{4\lambda^2}{\Gamma_\pi}\cdot\left(\frac{1}{\Gamma_\text{t}}+\frac{1}{\Gamma_\sigma}\right).
\label{eq1}
\end{equation}
In the tunneling range ($\Gamma_\sigma\gg\Gamma_\text{t}$) eq~\ref{eq1} may be further simplified to 
\begin{equation}
\frac{\Delta g}{g(0)}=\frac{4\lambda^2}{\Gamma_\pi \Gamma_\text{t}}.
\label{eq2}
\end{equation}
Thus, for similar $\Gamma_\text{t}$ and comparable $\lambda$, the IETS signal is controlled by $\Gamma_\pi$\@. 

With eqs~\ref{eq1} and \ref{eq2} all phonon-induced IET signatures and their evolution with varying junction conductance as reported here may be rationalized.
Moreover, the IET signal strengths of graphene phonons on other surfaces \cite{apl_91_122102,natphys_4_627,prl_104_036805,nl_11_2291,prl_114_245502,small_14_1703701,natcommun_6_7528,prb_95_161410} can be explained, as elaborated in the following.

A reduced coupling $\Gamma_\pi$ between graphene and the substrate corresponds to a longer lifetime of the $\pi$ state, which entails a stronger interaction with graphene phonons and, therefore, enhances the IET signals (Figure~\ref{fig1}d)\@.
This indicates that graphene on the Cs-intercalated samples is less hybridized with the metal than on Li-intercalated samples, which is consistent with the essentially vanishing moiré corrugation of graphene atop the Cs layer.
In the case of Ni intercalation the graphene $\pi$ states are strongly hybridized with Ni $3d$ bands \cite{prb_87_035420}, which in the simple model is reflected by a large $\Gamma_\pi$\@.
Therefore, the interaction of the $\pi$ state with graphene phonons is reduced and renders the inelastic channel inefficient. 
As a consequence, the current flows directly into the bulk of the metal substrate \textit{via} the elastic channel and the phonon signatures vanish from the $\text{d}I/\text{d}V$ spectra.

The variation of $\Gamma_\pi$ with increasing Li coverage is likely the cause for the evolution of $\Delta g/g(0)$ with $\Theta$ (Figure~\ref{fig1}e)\@.
A higher coverage of the Li intercalant progressively reduces the coupling to the metal substrate and, concomitantly, yields larger graphene phonon signals.
Besides the decoupling, charge transfer from Li to graphene leads to graphene doping, which may additionally promote IET signals owing to an increased density of states at the Fermi energy \cite{natphys_4_627,prl_104_036805,nl_11_2291,prl_114_245502}.
However, while Li and Cs provide similar doping at equal coverage \cite{prb_90_155428}, our experiments show that even on the densely packed Li film (global coverage $\approx$\,0.6\,ML), the IETS intensity is still well below that of the Cs-intercalated sample (Figure~\ref{fig1}d)\@. Consequently, the charge carrier density alone cannot adequately describe the graphene phonon excitation in IET, which is in disagreement with previous results \cite{natphys_4_627,prl_104_036805,nl_11_2291,prl_114_245502} and demonstrates the necessity of a comprehensive description.  

The developed model can likewise explain the extraordinarily high IET phonon signals observed from graphene on insulating and semiconducting surfaces \cite{apl_91_122102,natphys_4_627,prl_104_036805,nl_11_2291,prl_114_245502}, graphene blisters on Pt(111) \cite{natcommun_6_7528}, Ir(111) \cite{prb_95_161410} and from graphene bilayers \cite{small_14_1703701}.
In these cases, $\Gamma_\sigma$ is reduced, too, due to the low hybridization with substrate states at the Fermi level.
This scenario yields $\Gamma_\sigma \approx\Gamma_\text{t}\approx\Gamma_\pi $ (see eq~\ref{eq1}), which combines efficient inelastic transport with a small elastic current and leads to exceptionally large IET signals. The occasional absence of phonon spectroscopic signatures in $\text{d}I/\text{d}V$ spectra obtained for exfoliated graphene on SiO$_2$ \cite{prb_79_205411,natphys_7_245} may be explained by larger values of $\Gamma_\sigma$\@.
Indeed, the graphene--SiO$_2$ interface is characterized by charged impurities and single-electron charging effects giving rise to a substantial disorder potential \cite{natphys_7_245}.
The concomitant breaking of the graphene symmetry in weak-disorder systems leads to $\Gamma_\sigma$ dominating $\Gamma_\pi$ \cite{prb_91_121403}.
A similar argument was used previously to explain the absence of graphene phonon signals when the STM tip contacts the graphene sheet \cite{prb_95_161410}.
Consequently, $\Gamma_\pi$ and $\Gamma_\sigma$ act as control parameters that tune the efficiency of the inelastic tunneling pathway and, thus, the intensity of the phonon signals in IETS\@.

\begin{figure}
\includegraphics{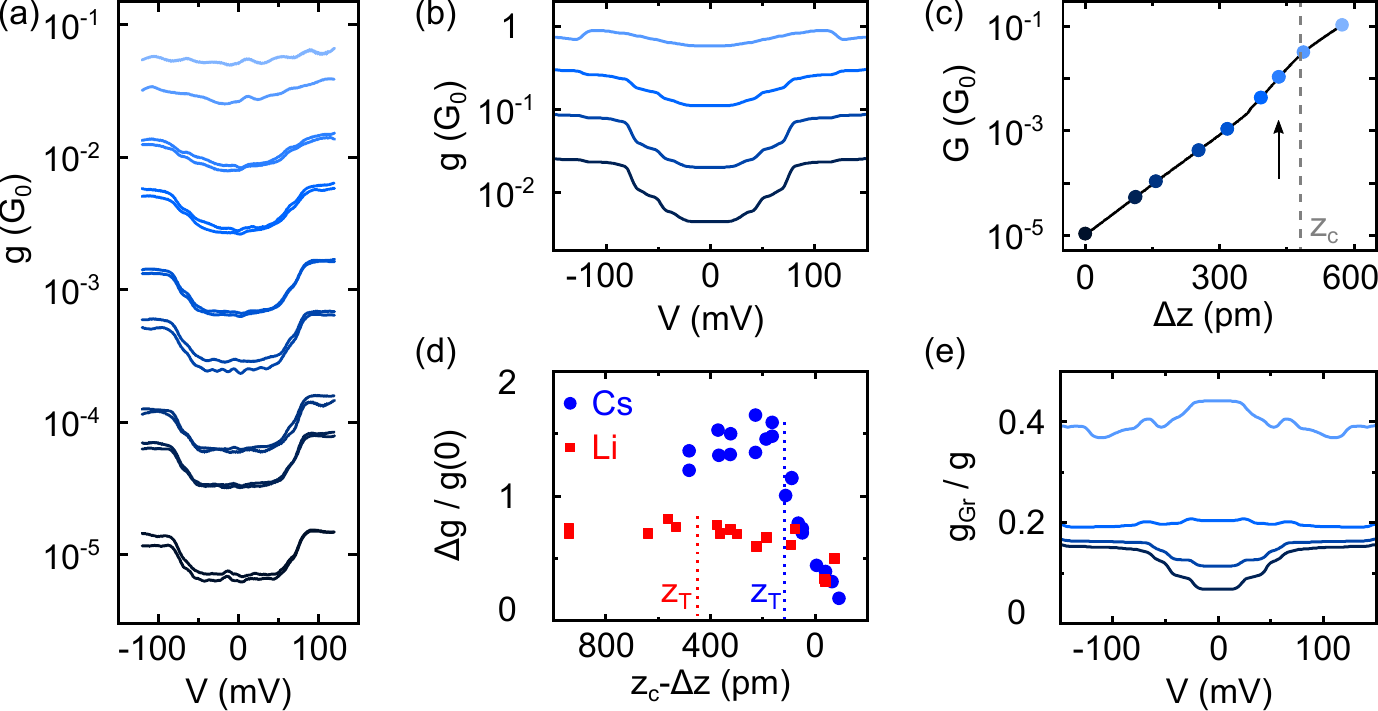}
\caption{(Color online) 
(a) Experimental $\text{d}I/\text{d}V$ ($g$) spectra of Cs-intercalated graphene for increasing (bottom to top) junction conductance showing the progressive quenching of the graphene phonon gap.
Closely spaced spectra reflect data acquired at tip approach and retraction.
(b) Simulated $\text{d}I/\text{d}V$ ($g$) spectra of Cs-intercalated graphene in the three-terminal model for junction conductances of 0.004 $G_0$, 0.02 $G_0$, 0.1 $G_0$, 0.6 $G_0$ (bottom to top).
(c) Junction conductance $G$ as a function of the tip displacement $\Delta z$ with $z_\text{c}$ the contact point (dashed line). 
$\Delta z=0\,\text{pm}$ is defined by $120\,\text{mV}$, $100\,\text{pA}$\@.
The arrow indicates the transition from tunneling ($\Delta z<365\,\text{pm}$) to contact ($\Delta z>486\,\text{pm}$).
Each dot marks the junction conductance at which spectra in (a) were acquired. 
(d) Phonon-induced relative changes, $\Delta g/g(0)$, for Cs (dots) and Li (squares) as a function of $z_c - \Delta z$ covering the range from tunneling to contact for both samples. The respective onsets z$_\text{T}$ of the transition from tunneling to contact are indicated by dotted lines. 
(e) Relative contribution ($g_{\text{Gr}}/g$) of the graphene terminal to the total calculated differential conductance $g$ for junction conductances as in (b).
}
\label{fig3}
\end{figure}

Not only the impact of the graphene--substrate coupling on the graphene phonon IETS signal strength may be described by the model.
From eqs~\ref{eq1} and \ref{eq2} the influence of the tip coupling $\Gamma_\text{t}$ may be examined as well.
Experimentally, the relative increase of $\text{d}I/\text{d}V$ due to phonon excitation is lowered with increasing junction conductance from tunneling to contact ranges, as shown for the Cs-intercalated sample in Figure~\ref{fig3}a. Figure~S4 of the Supporting Information shows the respective experimental data for Li. The experimentally observed trends for varying junction conductance are well captured by the simulations (Figure~\ref{fig3}b)\@.
The different transport ranges are best visualized in the evolution of the junction conductance, $G=I/V$, with $\Delta z$ (Figure~\ref{fig3}c) \cite{jpcm_20_223001,pccp_12_1022}.
The region of junction conductance indicated by the arrow in Figure~\ref{fig3}c  separates the tunneling ($\Delta z<365\,\text{pm}$) from the contact ($\Delta z>486\,\text{pm}$) range.
The displacement for contact formation, $z_\text{c}$, is defined by the intersection of exponential fits to conductance variations in the transition and contact ranges \cite{jpcm_20_223001,pccp_12_1022}.
Similar evolutions of the conductance were reported for graphene on Ru(0001) \cite{prl_105_236101}. 
At each junction conductance marked by dots in Figure~\ref{fig3}c the feedback loop was deactivated and an IET spectrum acquired.
The phonon-induced gap becomes shallower with increasing $G$\@.
At contact (topmost data sets in Figure~\ref{fig3}a) the IET signatures of graphene phonons have essentially disappeared.
This observation is in agreement with previous findings for graphene wrinkles \cite{prb_95_161410}.
Figure \ref{fig3}d summarizes the evolution of $\Delta g/g(0)$ for Cs and Li intercalants. 
In the whole conductance range from tunneling and transition to contact the sample intercalated by Cs produces larger IETS signals than the one intercalated by Li. 
Therefore, the difference between the intercalants cannot be rationalized in terms of a variation in the tip-sample distance alone.
Rather, it is indeed caused by different graphene--substrate couplings.

The close inspection of Figure~\ref{fig3}d reveals that the quenching of $\Delta g/g(0)$ is approximately twice as strong for Cs as for Li. According to eq~\ref{eq1}, different evolutions of $\Delta g/g(0)$ with the  tip-surface distance can be traced to the distance dependence of the three coupling constants $\Gamma_\sigma, \Gamma_\pi$, which are likely to depend on the intercalant, and $\Gamma_\text{t}$. For instance, $\Gamma_\sigma$ and $\Gamma_\pi$ may be reduced if graphene is locally detached from the surface due to the proximity of the tip. In previous contact experiments reported for graphene on Ru(0001) \cite{prl_105_236101} and on Ir(111) \cite{njp_16_053036} such elastic lifting of graphene was inferred from the gradual transition from tunneling to contact ranges in conductance-versus-distance data. However, as rationalized below, the observed quenching of $\Delta g/g(0)$ upon tip approach indicates the dominating role of $\Gamma_\text{t}$. First, $\Gamma_\text{t}$ increases with tip approach due to the increased van der Waals interaction between tip and graphene \cite{njp_16_053036,prb_90_075426}. Second, according to the model (Figure~\ref{fig2}b) a larger $\Gamma_\text{t}$ enhances both the elastic and inelastic transport channel.
Since $\Gamma_\sigma>\Gamma_\pi\gg\hbar\Omega$ the elastic channel is dominant and the ratio of inelastic and elastic currents decreases, and so does $\Delta g/g(0)$ (eq~\ref{eq1})\@.
An additional effect leading to the enhancement of both tunneling channels is the gradual lifting of the momentum conservation due to the local symmetry breaking by the approaching tip \cite{prb_91_125442}.
Indeed, the contribution of the graphene terminal to the total conductance rapidly rises (Figure~\ref{fig3}e)\@.
Close to contact many phonon modes from different regions of the surface Brillouin zone may contribute \cite{prb_95_161410}, which lowers the resolution of distinct phonon signatures in the IETS\@.
 
In conclusion, intercalation of graphene on a metal surface by Li and Cs leads to strong graphene phonon signatures in IETS with an STM.  Their signal strength can be tuned by the intercalant coverage as well as by the tip--surface separation ranging from tunneling to contact distances.  These experimental observations have sparked the comprehensive understanding of graphene phonon excitation in IETS on the basis of a three-terminal description.  The model calculations show how the electronic (covalent) coupling of graphene $\sigma$ and $\pi$ states with adjacent electrodes -- tip and sample -- regulates the current branching across the tunneling junction into elastic and inelastic transport channels.  We anticipate the general applicability of the proposed model to other two-dimensional materials, which currently attract substantial interest.

\begin{acknowledgement}
Financial support by the Deutsche Forschungsgemeinschaft through Grants No.~KR $2912/10-1$, KR $2912/12-1$, the Danish National Research Foundation through Project DNRF$103$ and by Villum Fonden through Grant No.~$00013340$ is acknowledged.
\end{acknowledgement}

\begin{suppinfo}
The Supporting Information is available free of charge on the ACS Publications website at DOI: [hyperlink DOI] \\
Details on sample preparation, Li spectroscopic data, IETS spatial variation, DFT modeling, NEGF calculation, contributing phonon modes
\end{suppinfo}

%\bibliography{ref}

\providecommand{\latin}[1]{#1}
\makeatletter
\providecommand{\doi}
  {\begingroup\let\do\@makeother\dospecials
  \catcode`\{=1 \catcode`\}=2 \doi@aux}
\providecommand{\doi@aux}[1]{\endgroup\texttt{#1}}
\makeatother
\providecommand*\mcitethebibliography{\thebibliography}
\csname @ifundefined\endcsname{endmcitethebibliography}
  {\let\endmcitethebibliography\endthebibliography}{}

\end{document}